\documentclass[twocolumn,prl,showpacs,preprintnumbers,amsmath,amssymb]{revtex4}

\usepackage{graphicx}% Include figure files
\usepackage{dcolumn}% Align table columns on decimal point
\usepackage{bm}% bold math

\begin{document}
\title{Thermodynamic properties of the coupled dimer system
Cu$_2$(C$_5$H$_{12}$N$_2$)$_2$Cl$_4$}
\author{S.\ Capponi} \affiliation{Laboratoire de
Physique Th\'eorique, Universit\'e Paul Sabatier, CNRS, 31062 Toulouse, France.}
\author{D.~Poilblanc}  \affiliation{Laboratoire de
Physique Th\'eorique, Universit\'e Paul Sabatier, CNRS, 31062 Toulouse, France.}

\date{\today}

\begin{abstract}

We re-examine the thermodynamic properties of the coupled dimer system
Cu$_2$(C$_5$H$_{12}$N$_2$)$_2$Cl$_4$ under magnetic field
in the light of recent  NMR experiments [Cl\'emancey {\it et al.}, Phys. Rev. Lett. {\bf 97}, 167204 (2006)]
suggesting the existence of a finite Dzyaloshinskii-Moriya interaction.
We show that including such a spin anisotropy
greatly improves the fit of the magnetization curve and
gives the correct trend of the insofar unexplained anomalous behavior of the
specific heat in magnetic field at low temperature.

\end{abstract}

\maketitle

The molecular solid Cu$_2$(C$_5$H$_{12}$N$_2$)$_2$Cl$_4$ is of
particular interest due to the formation of dimers between the two
copper (S=1/2) spins of each molecule. It was long believed that
the coupling between those dimers were quasi-one dimensional hence
realizing the physics of a two-leg spin ladder~\cite{Chaboussant}.
Later, it was emphasized that the magnetic coupling between
molecules might be more three-dimensional (3D)~\cite{INS-3D}. In a
recent Letter, Cl\'emancey {\it et al.}~\cite{ref1} proposed that
including Dzyaloshinskii-Moriya (DM) interactions is essential to
explain its experimental behavior under magnetic field. From
simple symmetry considerations, the lack of inversion centers at
the middle of the dimers enables {\it a priori} the existence of
such a DM term which can lead to experimentally observable effects
such as the appearance of a uniform transverse
magnetization~\cite{ref2} as shown by T=0 numerical calculations.
However, a theoretical investigation of physical properties at
{\it finite T} in the presence of such a DM interaction has not
been done sofar.

Following~\cite{ref1} we consider an anisotropic spin ladder
 under a
magnetic field $h$ along $z$ with a staggered DM term  along $y$,
\begin{eqnarray*}
{\cal H} &=& J_\perp \sum_i {\bf S}_{i,1} \cdot  {\bf S}_{i,2} +
J_\parallel  \sum_i ({\bf S}_{i,1} \cdot  {\bf S}_{i+1,1} + {\bf S}_{i,2} \cdot {\bf S}_{i+1,2}) \\
&+& D \sum_i (-1)^i \hat{\bf y} \cdot ({\bf S}_{i,1} \times {\bf S}_{i,2})
+ g \mu_B h \sum_i (S_{i,1}^z + S_{i,2}^z),
\end{eqnarray*}
where the last term corresponds to the field Zeeman energy.
Our aim here is to show that the thermodynamic properties
of the above-mentioned molecular system can be partly explained
within this model without invoking more complicated
three-dimensional couplings.

Specific heat  measurements done on the same compound could not be
satisfactorily explained with a SU(2)-invariant Heisenberg
interaction i.e. $D=0$~\cite{Calemczuk99,Hagiwara00,Batchelor}.
Indeed, despite a fairly good agreement at low fields, a large
discrepancy occurs for fields around $10$~T for which a strong
anomalous low-temperature peak around $1$~K is seen experimentally
and not reproduced theoretically. In contrast, we show here that
the inclusion of the finite DM interaction of~\cite{ref1} provides
a more quantitative agreement with the experimental data. We take
parameters of previous literature, $J_\perp=13.1$~K,
$J_\parallel=J_\perp/5$ and $g=2.1$, and use Exact
Diagonalisations (ED) to compute the magnetization curve and the
specific heat $C_{mag}(T,h)$. Interestingly, a finite $D$ strongly
reduces finite-size effects so that accurate estimates can be
obtained on 16-site systems (see later).

\begin{figure}[!ht]
\begin{center}
\includegraphics[width=0.49\textwidth]{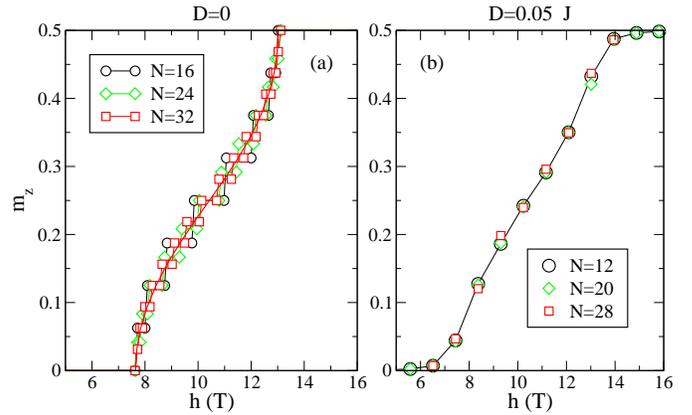}
\end{center}
\caption{(Color online)
Uniform magnetization per site $m_z$ vs magnetic field $h$ (in Tesla) for various cluster sizes.
(a) $D=0$~: $m_z$ exhibits finite-size plateaus at rational values
multiple of $1/N$ that disappear in the thermodynamic limit. (b)
Finite $D$~: we obtain continuous curves (since $S_z$ is not conserved
any more) which show very small
finite-size effects. The square
root singularity disappears and the magnetization
curves become smooth.}
\label{fig1}
\end{figure}

\begin{figure}[!ht]
\begin{center}
\includegraphics[width=0.45\textwidth]{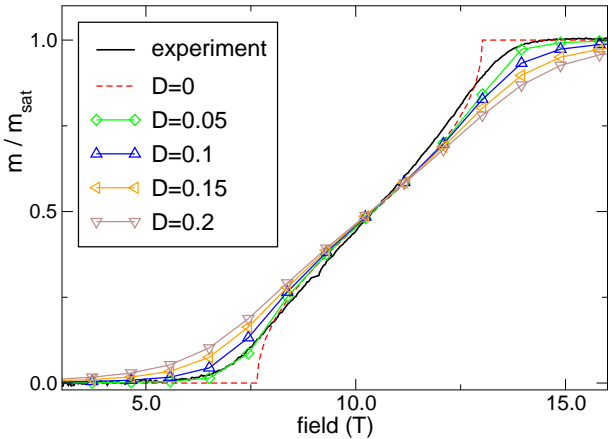}
\end{center}
\caption{(Color online)
Normalized magnetization along magnetic field $m_z$: experimental data
are taken
from Ref.~\onlinecite{Hayward96};
numerical results are presented for various $D$ (in units of $J_\perp$)
and $N=20$ spins (lines are guides to the eye).
In the specific case $D=0$, we plot an extrapolation (see text).
The best agreement with experiment is obtained for $D\simeq 0.05 J_\perp$.
}
\label{fig2}
\end{figure}

We start by computing the zero-temperature uniform magnetization along
the field $m_z$ for different sizes. As is shown on
Fig.~\ref{fig1}(a), when $D=0$, we recover the typical behavior
expected for an isotropic interaction with square root singularities~\cite{Affleck}
at the two critical fields $h_{C1}$ and $h_{C2}$. In that case, since
$S_z$ is conserved, it is sufficient to compute the ground-state
energy in all $S_z$ sectors (in the absence of any magnetic field) and
then perform a Legendre transform to get $m_z$. As a side remark, we
can mimic the thermodynamic limit by drawing a line connecting the
middles of the finite-size plateaus of the largest available system
($N=32$ sites). In contrast, for any finite $D$ (Fig.~\ref{fig1}(b)),
since $S_z$ is no longer a good quantum number, the magnetization has
to be computed for each $h$ value, which makes the problem harder.
Fortunately, as shown on Fig.~\ref{fig1}(b), the absence of $SU(2)$
symmetry makes the finite-size effects almost negligible for any
finite $D$. Moreover, we observe that the absence of singularities at
$h_{C1}$ and $h_{C2}$ is well reproduced by any finite $D$ as noticed
in Ref.~\onlinecite{ref2}.

Prior to the calculation of $C(T,h)$, we estimate $D$ from a fit of the
experimental magnetization curve~\cite{Hayward96} at low temperature. On  Fig.~\ref{fig2}, we plot
the numerical uniform magnetization for various $D$. By  comparing to experimental data,
we find that $D\simeq 0.05 J_\perp$ gives an excellent fit with no
other adjustable parameter. Remarkably, our estimate of $D$ is exactly the
numerical
value obtained in Ref.~\onlinecite{ref1}.

\begin{figure}[!ht]
\begin{center}
\includegraphics[width=0.5\textwidth]{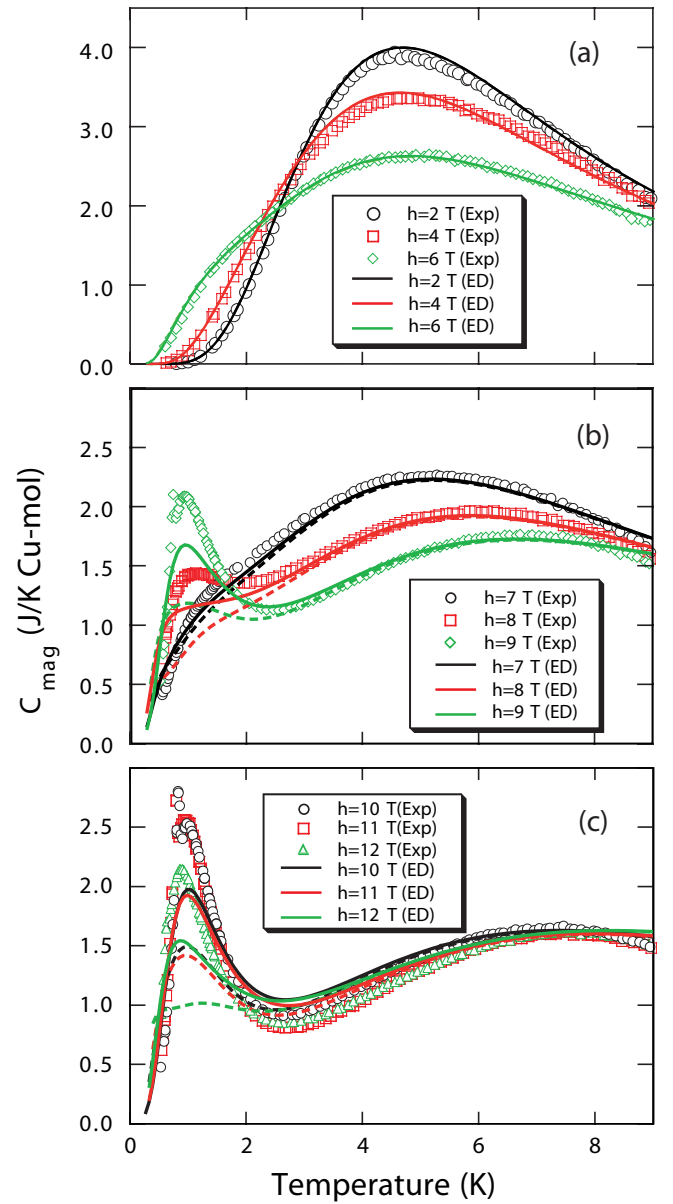}
\end{center}
\caption{(Color online)
Magnetic specific heat vs temperature for various magnetic fields $h$ (in Tesla)~: experimental data (symbols) are taken
from Ref.~\onlinecite{Hagiwara00}; exact numerical results are presented for $D=0$
(dashed lines, obtained with $20$ spins) and $D=0.05J_\perp$
(full lines, obtained with $16$ spins) showing a better agreement.
}
\label{fig3}
\end{figure}

We show on Fig.~\ref{fig3} the main results of our specific heat calculations and
compare it to the experimental data of Ref.~\onlinecite{Hagiwara00}.
First, at low magnetic field up to 6~T ($h< h_{c1}$), $D$ has
almost no effect due to the very large spin gap (the $D=0$ and $D=0.05 J_\perp$ curves are indistinguishable in Fig.\ref{fig3}(a))
 and the agreement is very good.
However, above this magnetic field, we observe, in addition to the
large broad maximum around 6~K, another peak around 1 or 2~K. In our results, this low-temperature
peak has the largest magnitude around  $h \sim 10$-$11$~T, as seen in experiment.
By taking our previous value, $D/J_\perp=0.05$, we clearly observe a significant increase of its intensity
that translates into a better comparison with experiments.

\begin{figure}[!ht]
\begin{center}
\includegraphics[width=0.5\textwidth]{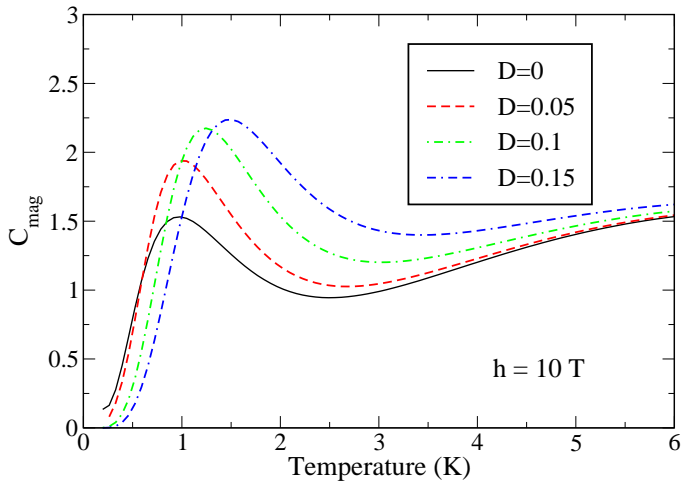}
\end{center}
\caption{(Color online) Magnetic specific heat for $h=10$~T
obtained numerically for various $D/J_\perp=0$, 0.05, 0.1 and 0.15
 with $16$ spins. As $D$ increases, we observe
that both the low-temperature peak intensity and position
increase.} \label{fig4}
\end{figure}

It is of interest to see whether a larger $D$ value could give rise to a better agreement for the heat capacity (at the
price of worsening the magnetization fit). However,
as indicated on Fig.~\ref{fig4}, we observe that increasing the DM interaction has two important
effects~: first, it leads to an increase of the height of the
low-temperature peak but, secondly, to a shift of its position in
temperature with little influence on the second (broad) maximum.
As a consequence, the minimum between the two peaks increases too
with $D$.

Therefore, a small DM term can  partly account for the anomalous
experimental height of the low-temperature peak. Note that small sources of
discrepancy between theory and experiment
should still remain like the experimental uncertainty
(due to the subtraction of the phonon contribution), the deviation
of the compound from a real ladder system, the small finite size
effects in our calculation, etc...

\begin{figure}[!ht]
\begin{center}
\includegraphics[width=0.45\textwidth]{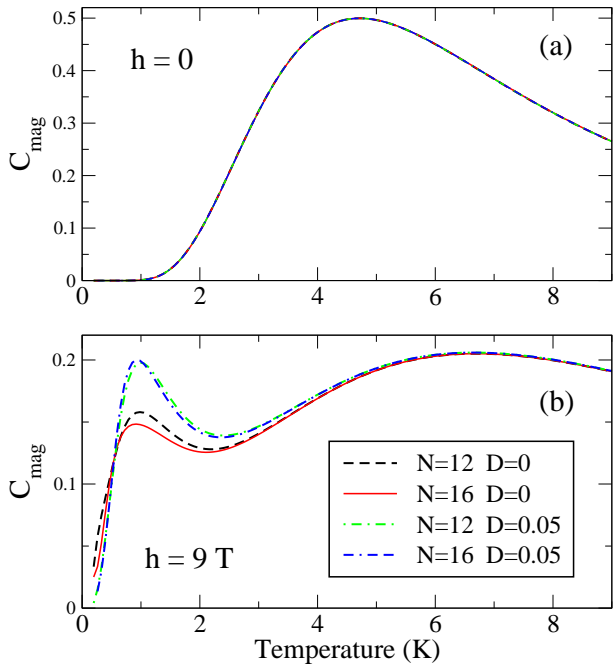}
\end{center}
\caption{(Color online) Magnetic specific heat for $h=0$ (a) and
$h=9$~T (b) obtained numerically for $D/J_\perp=0$ (full and dashed lines)
and 0.05 (dot-dashed lines) with $N=12$ and $N=16$ spins. Note that in
(a) all four curves are indistinguishable.} \label{fig5}
\end{figure}

Lastly, we would like to discuss briefly the finite size effects
at finite T on e.g. $C_{\rm mag}(T)$ by comparing data obtained on
clusters with 12 and 16 spins. As shown in Fig.~\ref{fig5}(a), in
the absence of a magnetic field, finite size effects are invisible
on the scale of the plot. Moreover, the specific heat is almost
unsensitive to a small finite DM term. This contrasts to the case of intermediate 
$h$ where, as shown above on Fig.~\ref{fig4}, the data are greatly sensitive to
the value of $D$. In addition, Fig.~\ref{fig5}(b) clearly shows 
that, fortunately, finite size effects are reduced when $D=0.05
J_\perp$ (compared to $D=0$), the two curves for $N=12$ and $N=16$
being almost superposed. This provides a great deal of confidence
on our previous analysis.

To conclude, we have shown that the proposal of
Ref.~\onlinecite{ref1} of a finite (small) DM interaction gives the
correct trend of the insofar unexplained anomalous behavior of the
specific heat in magnetic field of
Cu$_2$(C$_5$H$_{12}$N$_2$)$_2$Cl$_4$. From magnetization measurements,
our considerations provide
an estimate of $D$ around $0.05 J_\perp$ in agreement with the
value extracted from other recent measurements~\cite{ref1}.

\begin{acknowledgments}
We thank S.~Miyahara and F.~Mila for fruitful discussions.
\end{acknowledgments}

\end{document}